\def\la{\hbox{{\lower -2.5pt\hbox{$<$}}\hskip -8pt\raise
-2.5pt\hbox{$\sim$}}}
\def\ga{\hbox{{\lower -2.5pt\hbox{$>$}}\hskip -8pt\raise
-2.5pt\hbox{$\sim$}}}
\def\ltsima{$\; \buildrel < \over \sim \;$}
\def\simlt{\lower.5ex\hbox{\ltsima}}
\def\gtsima{$\; \buildrel > \over \sim \;$}
\def\simgt{\lower.5ex\hbox{\gtsima}}

\documentstyle[epsfig]{elsart}

\begin{document}
\begin{frontmatter}
\title{Non-linear shock acceleration in the presence of seed particles}
\author[Fermi,infn]{Pasquale Blasi\thanksref{corr}} 
\address[Fermi]{INAF/Osservatorio Astrofisico di Arcetri,\\
Largo E. Fermi, 5 I-50125 Firenze (Italy)}
\address[infn]{INFN, Sezione di Firenze}
\thanks[corr]{E-mail: blasi@arcetri.astro.it}

\begin{abstract}
Particles crossing repeatedly the surface of a shock wave can be energized 
by first order Fermi acceleration. The linear theory is successful in 
describing the acceleration process as long as the pressure of the accelerated
particles remains negligible compared to the kinetic pressure of the 
incoming gas (the so-called test particle approximation). When this condition 
is no longer fulfilled, the shock is modified by the pressure of the 
accelerated particles in a nonlinear way, namely the
spectrum of accelerated particles and the shock structure determine each
other. In this paper we present the first description of the nonlinear 
regime of shock acceleration when the shock propagates in a medium
where seed particles are already present. 
This case may apply for instance to supernova shocks propagating into the 
interstellar medium, where cosmic rays are in equipartition with the
gas pressure. We find that the appearance of multiple solutions, previously
found in alternative descriptions of the nonlinear regime, occurs also for 
the case of reacceleration of seed particles. 
Moreover, for parameters of concern for supernova shocks, the shock
is likely to turn nonlinear mainly due to the presence of the pre-existing
cosmic rays, rather than due to the acceleration of new particles from the 
thermal pool. We investigate here the onset of the nonlinear regime for
the three following cases: 1) seed particles in equipartition with the
gas pressure; 2) particles accelerated from the thermal pool; 3)
combination of 1) and 2).

\end{abstract}

\end{frontmatter}

\section{Introduction}
Diffusive acceleration at newtonian shock fronts is an extensively studied 
phenomenon. Detailed discussions of the current status of the investigations 
can be found in some recent excellent reviews \cite{drury83,be87,bk88,je91}. 
While much is by now well understood, some issues are still subjects
of much debate, for the theoretical and phenomenological implications that 
they may have. The most important of these is the backreaction of the 
accelerated particles on the shock: the violation of the 
{\it test particle approximation} occurs when the acceleration process 
becomes sufficiently efficient to generate pressures of the accelerated
particles which are comparable with the incoming gas kinetic pressure.
Both the spectrum of the particles and the structure of the shock
are changed by this phenomenon, which is therefore intrinsically 
nonlinear.

At present there are three viable approaches to account for the backreaction 
of the particles upon the shock: one is based on the ever-improving numerical
simulations \cite{je91,bell87,elli90,ebj95,ebj96,kj97,jones02} that allow 
a self-consistent treatment of several effects. 
The second approach is a {\it fluid} approach, and treats cosmic rays as
a relativistic second fluid. This class of models was proposed and discussed
in \cite{dr_v80,dr_v81,dr_ax_su82,ax_l_mk82,ddv94}. These models allow one
to obtain the thermodynamics of the modified shocks, but do not provide
information about the spectrum of accelerated particles.
 
The third approach is analytical and may be very helpful to understand 
the physics of the nonlinear effects in a way that sometimes is difficult
to achieve through simulations, due to their intrinsic complexity. 

In Ref. \cite{blandford80} a perturbative approach was adopted, in which 
the pressure of accelerated particles was treated as a small perturbation.
By construction this method provides an answer only for weakly modified
shocks.

An alternative approach was proposed in 
\cite{eich84a,eich84b,eich85,elleich85},
based on the assumption that the diffusion of the particles
is sufficiently energy dependent that different parts of the fluid are
affected by particles with different average energies. The way the calculations
are carried out implies a sort of separate solution of the transport equation
for subrelativistic and relativistic particles, so that the two spectra must 
be somehow connected at $p\sim mc$ {\it a posteriori}. 

Recently, in \cite{berezhko94,berezhko95,berezhko96}, the effects of the 
non-linear backreaction of accelerated particles on the maximum achievable
energy were investigated, together with the effects of geometry. The 
maximum energy of the particles accelerated in supernova remnants in
the presence of large acceleration efficiencies was also studied in 
\cite{ptuskin1,ptuskin2}.

The need for a {\it practical} solution of the acceleration problem in the 
non-linear regime was recognized in \cite{simple}, where a simple analytical
broken-power-law approximation of the non-linear spectra was presented. 

Recently, some promising analytical solutions of the problem of non-linear 
shock acceleration have appeared in the literature 
\cite{malkov1,malkov2,blasi02}. These solutions
seem to avoid many of the limitations of previous approaches. 


Numerical simulations have been instrumental to identify the dramatic
effects of the particles backreaction: they showed that even when the fraction 
of particles injected from the thermal gas is relatively small, the energy 
channelled into these few particles can be an appreciable part of the kinetic 
energy of the unshocked fluid, making the test particle approach unsuitable. 
The most visible effects of the backreaction of the accelerated particles on 
the shock appear in the spectrum of the accelerated particles, which shows
a peculiar flattening at the highest energies. The analytical approaches
reproduce well the basic features arising from nonlinear effects in
shock acceleration. 

While several calculations exist of the nonlinear effects in the shock 
acceleration of quasi-monochromatic particles injected at a shock surface,
there is no description at present of how these effects appear, if they
do, when the shock propagates in a medium where (pre)accelerated particles 
already exist. The linear theory of this phenomenon was developed by Bell 
\cite{bell78}, but has never been generalized to its nonlinear extension.
In fact, the backreaction can severely affect the process of re-energization 
of preaccelerated particles: Bell already showed that for strong shocks the 
energy content of a region where cosmic rays were present could be easily 
enhanced by a factor $\sim 10$ at each shock passage, so that equipartition 
could be readily reached. In these conditions the backreaction of the 
accelerated particles should be expected.

We report here on the first analytical treatment of the shock acceleration
in the presence of seed nonthermal particles, with the inclusion of their 
nonlinear backreaction on the shock. Our approach is a generalization
of the analytical method introduced in \cite{blasi02} to describe the 
nonlinear shock acceleration with monochromatic injection of quasi-thermal
particles. In fact, we present here also a general calculation that accounts
for both thermal particles and seed nonthermal particles. This case may be of
interest for the study of supernova shocks propagating through the 
interstellar medium (ISM) where pressure balance exists between gas and 
cosmic rays.

Nonlinear effects in shock acceleration of thermal particles result in 
the appearance of multiple solutions in certain regions of the parameter
space. This behaviour resembles that of critical systems, with a bifurcation
occurring when some threshold is reached. In the case of shock acceleration, 
it is not easy to find a way of discriminating among the multiple 
solutions when they appear. Neverthless, in \cite{mond}, a two fluid approach
has been used to demonstrate that when three solutions appear, the one with 
intermediate efficiency for particle acceleration is unstable to corrugations
in the shock structure and emission of acustic waves. Plausibility arguments 
may be put forward to justify that the system made of the shock plus the 
accelerated particles may sit at the critical point, but the author is not 
aware of any real proof that this is what happens. The physical parameters 
that play a role in this approach 
to criticality are the maximum momentum achievable by the particles in the 
acceleration process, the Mach number of the shock, and the injection 
efficiency, namely the fraction of thermal particles crossing the shock that 
are accelerated to nonthermal energies.
The last of them, the injection efficiency, hides a crucial
physics problem by itself, and may play an important role in establishing 
the level of shock modification. This efficiency parameter in reality is 
defined by the microphysics of the shock and should not be a free parameter 
of the problem. Unfortunately, our poor knowledge of such microphysics, in 
particular for collisionless shocks, does not allow us to establish a
clear and universal connection between the injection efficiency and the
macroscopic shock properties.

The paper is structured as follows: in \S \ref{sec:nonlin} we describe 
the effect of non linearity on shock acceleration and our mathematical
approach to describe it. In particular we generalize previous calculations 
to the case in which seed particles exist in the region where the shock is 
propagating; in \S \ref{sec:gasdyn} we describe the gas dynamics in the
presence of a non-negligible pressure of accelerated particles.
In \S \ref{sec:results} we describe our results, with particular attention
for the onset of the particle backreaction and for the appearance of 
multiple solutions. We conclude in \S \ref{sec:conclusions}.

\section{Non linear shock acceleration}\label{sec:nonlin}

In this section we solve the diffusion-convection equation for the cosmic
rays in the most general case in which particles are injected according to
some function $Q(x,p)$ and the shock propagates within a region where some 
cosmic ray distribution exists, that we denote as $f_\infty(p)$, spatially
homogeneous before the shock crossing (upstream). 

For simplicity we limit ourselves to the case of one-dimensional shocks,
but the introduction of different geometrical effects is relatively simple, 
and in fact many of our conclusions should not be affected by geometry. 

The equation that describes the diffusive transport of particles 
in one dimension is
\begin{equation}
\frac{\partial}{\partial x}
\left[ D  \frac{\partial}{\partial x} f(x,p) \right] - 
u  \frac{\partial f (x,p)}{\partial x} + \frac{1}{3} 
 \frac{d u}{d x}~p~\frac{\partial f(x,p)}{\partial p} + 
Q(x,p) = 0,
\label{eq:trans}
\end{equation}
where we assumed stationarity ($\partial f/\partial t=0$). The $x$ axis is
oriented from upstream to downstream, as in fig. 1. The presence of 
pre-existing cosmic rays is introduced as a boundary condition at upstream 
infinity, by requiring that $f(x=-\infty,p)=f_\infty (p)$. One should keep 
in mind that the common picture of a fluid whose speed is constant ($u_1$) 
until it hits the shock surface is not appropriate for modified shocks. 
In fact, in this case, the pressure of the accelerated particles may become 
large enough to slow down the fluid before it crosses the shock surface. 
Therefore in general at upstream infinity the gas flows at
speed $u_0$, different from $u_1$ (fluid speed immediately upstream of the 
shock). The two quantities are approximately 
equal only when the accelerated particles do not have dynamical relevance. The
injection term is taken in the form $Q(x,p)=Q_0(p) \delta(x)$.    
\begin{figure}[thb]
 \begin{center}
  \mbox{\epsfig{file=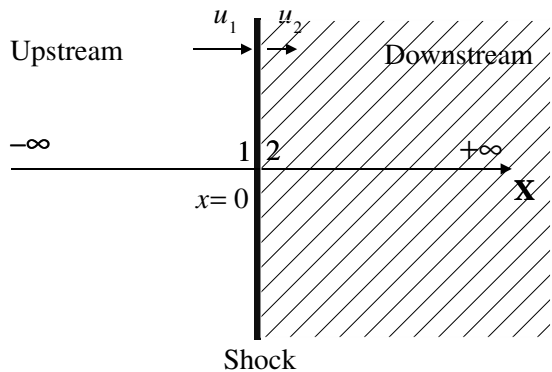,width=13.cm}}
  \caption{Schematic view of the shock region.}
 \end{center}
\end{figure}
As a first step, we integrate eq. \ref{eq:trans} around $x=0$, from $x=0^-$ to
$x=0^+$, denoted in fig. 1 as points ``1'' and ``2'' respectively, so that the
following equation can be written:
\begin{equation}
\left[ D \frac{\partial f}{\partial x}\right]_2 -
\left[ D \frac{\partial f}{\partial x}\right]_1 +
\frac{1}{3} p \frac{d f_0}{d p} (u_2 - u_1) + Q_0(p)= 0,
\end{equation}
where $u_1$ ($u_2$) is the fluid speed immediately upstream (downstream) 
of the shock and $f_0$ is the particle distribution function at the shock
location.
By requiring that the distribution function downstream is independent of
the spatial coordinate (homogeneity), we obtain
$\left[ D \frac{\partial f}{\partial x}\right]_2=0$, so that the boundary 
condition at the shock can be rewritten as
\begin{equation}
\left[ D \frac{\partial f}{\partial x}\right]_1 =
\frac{1}{3} p \frac{d f_0}{d p} (u_2 - u_1) + Q_0(p).
\label{eq:boundaryshock}
\end{equation}
We can now perform the integration of eq. (\ref{eq:trans}) from $x=-\infty$ to
$x=0^-$ (point ``1''), in order to take into account the boundary condition at 
upstream infinity. Using eq. (\ref{eq:boundaryshock}) we obtain
\begin{equation}
\frac{1}{3} p \frac{d f_0}{d p} (u_2 - u_1) - u_1 f_0 + u_0 f_\infty + Q_0(p)+
\int_{-\infty}^{0^-} dx f \frac{d u}{d x} + \frac{1}{3}\int_{-\infty}^{0^-} 
dx  \frac{d u}{d x} p \frac{\partial f}{\partial p} = 0.
\label{eq:step}
\end{equation}
We can now introduce a quantity $u_p$ defined as
\begin{equation}
u_p = u_1 - \frac{1}{f_0} \int_{-\infty}^{0^-} dx \frac{d u}{d x} f(x,p),
\label{eq:up}
\end{equation}
whose physical meaning is instrumental to understand the nonlinear 
backreaction of particles. The function $u_p$ is the average fluid 
velocity experienced by particles with momentum $p$ while diffusing 
upstream away from the shock surface. In other words, the effect of the 
average is that, instead of a constant speed $u_1$ upstream, a particle 
with momentum $p$ experiences a spatially variable speed, due to the 
pressure of the accelerated particles that is slowing down the fluid. Since 
the diffusion coefficient is in general $p$-dependent, particles with 
different energies {\it feel} a different compression coefficient, higher 
at higher energies if, as expected, the diffusion coefficient is an 
increasing function of momentum.

The role of $u_p$ can also be explained as follows: the distribution function
$f(x,p)$ at a distance $x$ from the shock surface can be written as
\cite{malkov2}
$$f(x,p)=f_0(p) \exp \left[ \frac{q(p)}{3 D(p)}
\int_{0}^x dx' u(x') \right],$$
where $q(p)=-d\ln f_0(p)/d \ln p$ is the local slope of $f_0(p)$ and 
the diffusion coefficient $D(p)$ has been assumed independent of the
location $x$. 
In first approximation, we can assume that the exponential factor remains
important when it is of order unity, namely when its argument is much 
less than unity. We can therefore introduce a distance $x_p$, which is 
the distance at which the exponential equals one. This means that 
$$u_p \approx u_1 - \int_{-x_p}^{0-} dx \frac{du}{dx} \approx 
u_1-\left[u_1-u(x_p)\right]\approx u(x_p),$$
so that the speed $u_p$ can be interpreted as the fluid speed at the 
point $x_p$ where the particles with momentum $p$ reverse their motion in 
the upstream fluid and return to the shock. Note that the diffusion 
coefficient enters the calculation of the distance $x_p$ but does not 
enter directly the calculation of $u_p$. In other words, different
diffusion coefficients may move the point where the fluid speed is 
$u_p$ closer to or farther from the shock surface, but do not affect 
the value of $u_p$. This approach is similar to that introduced in 
\cite{eich84a}.

With the introduction of $u_p$, eq. (\ref{eq:step}) becomes 
\begin{equation}
\frac{1}{3} p \frac{d f_0}{d p} (u_2 - u_p) - f_0 \left[u_p+\frac{1}{3} 
p \frac{du_p}{dp} \right] + u_0 f_\infty + Q_0(p) = 0 ,
\label{eq:step1}
\end{equation}
where we used the fact that
$$
p\frac{d}{dp}\int_{-\infty}^{0^-} dx \frac{du}{dx} f = 
p \left[\frac{df_0}{dp} (u_1-u_p) -f_0 \frac{du_p}{dp} \right].
$$
Eq. (\ref{eq:step1}) can be written in a way that resembles more the 
equation for shocks with no particles backreaction but in the presence 
of seed particles with distribution $f_\infty(p)$ and injection function 
$Q_0(p)$:
\begin{equation}
p \frac{d f_0}{d p} = -\frac{3}{u_p - u_2}\left\{ f_0 \left[ u_p + \frac{1}{3} 
p \frac{du_p}{dp} \right] - u_0 f_\infty -Q_0(p) \right\}.
\label{eq:transport}
\end{equation}
The solution of this equation can be written in the following implicit form:
$$
f_0(p) = f_0^{reacc} (p) + f_0^{inj}(p) =
$$
\begin{equation}
\int_{p_0}^{p} \frac{d{\bar p}}{{\bar p}} 
\frac{3 \left[u_0 f_\infty ({\bar p}) 
+ Q_0({\bar p})\right]}{u_{\bar p} - u_2} 
\exp\left\{-\int_{\bar p}^p \frac{dp'}{p'} 
\frac{3}{u_{p'} - u_2}\left[u_{p'}+\frac{1}{3}p' \frac{du_{p'}}{d p'}\right]
\right\}.
\label{eq:solut}
\end{equation}
In the case of monochromatic injection with momentum $p_{inj}$ at the shock 
surface, we can write
\begin{equation}
Q_0(p) = \frac{\eta n_{gas,1} u_1}{4\pi p_{inj}^2} \delta(p-p_{inj}),
\end{equation}  
where $n_{gas,1}$ is the gas density immediately upstream ($x=0^-$) and $\eta$ 
parametrizes the fraction of the particles crossing the shock which are 
going to take part in the acceleration process. The injection term in eq. 
(\ref{eq:solut}) becomes
$$
f_0^{inj} (p) = \int_{p_0}^{p} \frac{d{\bar p}}{{\bar p}} 
\frac{3 Q_0({\bar p})}{u_{\bar p} - u_2} \exp\left\{-\int_{\bar p}^p 
\frac{dp'}{p'} \frac{3}{u_{p'} - u_2}\left[u_{p'}+\frac{1}{3}p' 
\frac{du_{p'}}{d p'}\right]\right\}=
$$
\begin{equation}
\frac{3 R_{sub}}{R_{sub}-1} \frac{\eta n_{gas,1}}{4\pi p_{inj}^3} 
\exp\left\{-\int_{p_{inj}}^p 
\frac{dp'}{p'} \frac{3}{u_{p'} - u_2}\left[u_{p'}+\frac{1}{3}p' 
\frac{du_{p'}}{d p'}\right]\right\}.
\label{eq:inje}
\end{equation}
Here we introduced the two quantities $R_{sub}=u_1/u_2$ and $R_{tot}=u_0/u_2$, 
which are respectively the compression factor at the gas subshock and 
the total compression factor between upstream infinity and downstream. The 
two compression factors would be equal in the test particle approximation.
For a modified shock, $R_{tot}$ can attain values much larger than
$R_{sub}$ and more in general, much larger than $4$, which is the maximum value
achievable for an ordinary strong non-relativistic shock. The increase of the 
total compression factor compared with the prediction for an ordinary shock is 
responsible for the peculiar flattening of the spectra of accelerated 
particles that represents a feature of nonlinear effects in shock acceleration.
In terms of $R_{sub}$ and $R_{tot}$ the density immediately upstream is 
$n_{gas,1}=(\rho_0/m_p)R_{tot}/R_{sub}$.

In eq. (\ref{eq:inje}) we can introduce a dimensionless quantity 
$U(p)=u_p/u_0$ so that
\begin{equation}
f_0^{inj} (p) = \left(\frac{3 R_{sub}}{R_{tot} U(p) - 1}\right) 
\frac{\eta n_{gas,1}}{4\pi p_{inj}^3} \exp \left\{-\int_{p_{inj}}^p 
\frac{dp'}{p'} \frac{3R_{tot}U(p')}{R_{tot} U(p') - 1}\right\}.
\end{equation}

Introducing the same formalism also for the reacceleration term in eq. 
(\ref{eq:solut}), we obtain the general expression
$$
f_0(p) = \frac{3 R_{tot}}{R_{tot} U(p) - 1} 
\int_{p_0}^{p} \frac{d{\bar p}}{{\bar p}} f_\infty ({\bar p}) 
\exp\left\{-\int_{\bar p}^p \frac{dp'}{p'} 
\frac{3R_{tot} U(p')}{R_{tot} U(p') - 1}\right\} +
$$
\begin{equation}
\left(\frac{3 R_{sub}}{R_{tot} U(p) - 1}\right) 
\frac{\eta n_{gas,1}}{4\pi p_{inj}^3} \exp \left\{-\int_{p_{inj}}^p 
\frac{dp'}{p'} \frac{3R_{tot}U(p')}{R_{tot} U(p') - 1}\right\}.
\label{eq:laeffe}
\end{equation}

The solution of the problem is known if the velocity field $U(p)=u_p/u_0$ is known.
The nonlinearity of the problem reflects in the fact that $U(p)$ is in turn a function
of $f_0$ as it is clear from the definition of $u_p$. 
In order to solve the problem we need to write the equations for the thermodynamics
of the system including the gas, the reaccelerated cosmic rays, the cosmic rays 
accelerated from the thermal pool and the shock itself. We write and solve these 
equations in the next section.

\section{The gas dynamics of modified shocks with seed particles} \label{sec:gasdyn}

The velocity, density and thermodynamic properties of the fluid
can be determined by the mass and momentum conservation equations, with the 
inclusion of the pressure of the accelerated particles and of the preexisting
cosmic rays. We write these equations between a point far upstream 
($x=-\infty$), where the fluid velocity is $u_0$ and the density is 
$\rho_0=m n_{gas,0}$, and the point where the fluid upstream velocity is 
$u_p$ (density $\rho_p$).
The index $p$ denotes quantities measured at the point where the
fluid velocity is $u_p$, namely at the point $x_p$ that can be reached
only by particles with momentum $\geq p$.

The mass conservation implies:
\begin{equation}
\rho_0 u_0 = \rho_p u_p.
\label{eq:mass}
\end{equation}
Conservation of momentum reads:
\begin{equation}
\rho_0 u_0^2 + P_{g,0} + P_{CR,0} = \rho_p u_p^2 + P_{g,p} + P_{CR,p},
\label{eq:pressure}
\end{equation}
where $P_{g,0}$ and $P_{g,p}$ are the gas pressures at the points 
$x=-\infty$ and $x=x_p$ respectively, and $P_{CR,p}$ is the pressure
in accelerated particles at the point $x_p$ (we used the symbol $CR$
to mean {\it cosmic rays}, in the sense of {\it accelerated particles}).
The mass flow in accelerated particles has reasonably been neglected.

Our basic assumption, similar to that used in \cite{eich84a}, is that
the diffusion is $p$-dependent and more specifically that the diffusion
coefficient $D(p)$ is an increasing function of $p$. Therefore the typical
distance that a particle with momentum $p$ moves away from the shock is
approximately $\Delta x\sim D(p)/u_p$, larger for high energy particles
than for lower energy particles\footnote{For the cases of interest, $D(p)$
increases with $p$ faster than $u_p$ does, therefore $\Delta x$ is a
monotonically increasing function of $p$.}. As a consequence, at each 
given point $x_p$ only particles with momentum larger than $p$ are able 
to affect appreciably the fluid. Strictly speaking the validity of the 
assumption depends on how strongly the diffusion coefficient depends on 
the momentum $p$.

The cosmic ray pressure at $x_p$ is the sum of two terms: one is the pressure
contributed by the adiabatic compression of the cosmic rays at upstream 
infinity, and the second is the pressure of the particles accelerated or 
reaccelerated at the shock (${\tilde P}_{CR}(p)$) and able to reach the 
position $x_p$. Since only particles with momentum $\simgt p$ can reach 
the point $x=x_p$, we can write
$$
P_{CR,p} = P_{CR,0} \left(\frac{u_0}{u_p}\right)^{\gamma_{CR}} + 
{\tilde P}_{CR}(p) \simeq
$$
\begin{equation}
\simeq P_{CR,0}(p) \left(\frac{u_0}{u_p}\right)^{\gamma_{CR}} + 
\frac{4\pi}{3} \int_{p}^{p_{max}} dp p^3 v(p) f_0(p),
\label{eq:CR}
\end{equation}
where $v(p)$ is the velocity of particles with momentum $p$, $p_{max}$ 
is the maximum momentum achievable in the specific situation under 
investigation, and $\gamma_{CR}$ is the adiabatic index for the 
accelerated particles. 

Let us consider separately the case of a strongly modified and weakly
modified shock, in order to determine the best choice for $\gamma_{CR}$.
In the case of strongly modified shocks, we will show that most energy
is piled up in the region $p\sim p_{max}\gg m$, therefore in this case 
we can safely adopt $\gamma_{CR}=4/3$, appropriate for a relativistic
gas. For weakly modified shocks, the accelerated particles have an
approximately power law spectrum with a slope $\alpha$. It can be
shown that in this case $\gamma_{CR}=\alpha/3$, so that the relativistic
result $\gamma_{CR}=4/3$ still applies for $\alpha=4$ (strong shocks). 
For steeper spectra ($\alpha>4$) a larger adiabatic index should be 
adopted, but in those cases the solution is basically independent of 
the choice of $\gamma_{CR}$ because of the weak backreaction of the
particles. For the purpose of carrying out our numerical calculation 
we will therefore always take $\gamma_{CR}=4/3$. 

The pressure of cosmic rays at upstream infinity is simply
\begin{equation}
P_{CR,0}=\frac{4\pi}{3} \int_{p_{min}}^{p_{max}} dp p^3 v(p) f_\infty(p),
\end{equation}
where $p_{min}$ is some minimum momentum in the spectrum of seed particles.
For simplicity, we assume that $p_{min}=p_{inj}$, namely the minimum 
momentum of the seed particles coincides with the momentum at which 
particles are injected in the shock and are accelerated.

From eq. (\ref{eq:pressure}) we can see that there is a maximum distance, 
corresponding to the propagation of particles with momentum $p_{max}$ 
such that at larger distances the fluid is unaffected by the accelerated 
particles and $u_p=u_0$.

We will show later that for strongly modified shocks the integral in eq. 
(\ref{eq:CR}) is dominated by the region $p\sim p_{max}$. This improves
even more the validity of our approximation $P_{CR,p}=P_{CR}(>p)$.
This also suggests that different choices for the diffusion coefficient
$D(p)$ may affect the value of $p_{max}$, but at fixed $p_{max}$ the 
spectra of the accelerated particles should not change in a significant way.

Assuming an adiabatic compression of the gas in the upstream region, 
we can write
\begin{equation} 
P_{g,p}=P_{g,0} \left(\frac{\rho_p}{\rho_0}\right)^{\gamma_g}=
P_{g,0} \left(\frac{u_0}{u_p}\right)^{\gamma_g},
\label{eq:Pgas}
\end{equation}
where we used mass conservation, eq. (\ref{eq:mass}). The gas pressure 
far upstream is $P_{g,0}=\rho_0 u_0^2/(\gamma_g M_0^2)$, where $\gamma_g$ 
is the ratio of specific heats for the gas ($\gamma_g=5/3$ for an ideal 
gas) and $M_0$ is the Mach number of the fluid far upstream.

We introduce now a parameter $\xi_{CR}$ that quantifies the relative weight 
of the cosmic ray pressure at upstream infinity compared with the pressure
of the gas at the same location, $\xi_{CR}=P_{CR,0}/P_{g,0}$. Using this
parameter and the definition of the function $U(p)$, the equation for momentum 
conservation becomes
\begin{equation}
\frac{dU}{dp} \left[ 1 - \frac{\gamma_{CR}}{\gamma_g} \frac{\xi_{CR}}{M_0^2}
U^{-(\gamma_{CR}+1)} - \frac{1}{M_0^2} U^{-(\gamma_g+1)} \right] + 
\frac{1}{\rho_0 u_0^2} \frac{d{\tilde P}_{CR}}{dp} = 0.
\end{equation}
Using the definition of ${\tilde P}_{CR}$ and multiplying by $p$, this equation 
becomes
\begin{equation}
p\frac{dU}{dp} \left[ 1 - \frac{\gamma_{CR}}{\gamma_g} \frac{\xi_{CR}}{M_0^2}
U^{-(\gamma_{CR}+1)} - \frac{1}{M_0^2} U^{-(\gamma_g+1)} \right] = 
\frac{4\pi}{3 \rho_0 u_0^2} p^4 v(p) f_0(p),
\label{eq:eqtosolve}
\end{equation}
where $f_0$ depends on $U(p)$ as written in eq. (\ref{eq:laeffe}). Eq. 
(\ref{eq:eqtosolve}) is therefore an integral-differential nonlinear 
equation for $U(p)$. The solution of this equation also provides the 
spectrum of the accelerated particles.
 
The last missing piece is the connection between $R_{sub}$ and $R_{tot}$, the
two compression factors appearing in eq. (\ref{eq:solut}). The compression 
factor at the gas shock around $x=0$ can be written in terms of the Mach number 
$M_1$ of the gas immediately upstream through the well known expression
 
\begin{equation}
R_{sub} = \frac{(\gamma_g+1)M_1^2}{(\gamma_g-1)M_1^2 + 2}.
\end{equation}
On the other hand, if the upstream gas evolution is adiabatic, then the Mach
number at $x=0^-$ can be written in terms of the Mach number of the fluid at
upstream infinity $M_0$ as
$$
M_1^2 = M_0^2 \left( \frac{u_1}{u_0} \right)^{\gamma_g+1} =
M_0^2 \left( \frac{R_{sub}}{R_{tot}} \right)^{\gamma_g+1},
$$
so that from the expression for $R_{sub}$ we obtain
\begin{equation}
R_{tot} = M_0^{\frac{2}{\gamma_g+1}} \left[ 
\frac{(\gamma_g+1)R_{sub}^{\gamma_g} - (\gamma_g-1)R_{sub}^{\gamma_g+1}}{2}
\right]^{\frac{1}{\gamma_g+1}}.
\label{eq:Rsub_Rtot}
\end{equation}

Now that an expression between $R_{sub}$ and $R_{tot}$ has been found, eq.
(\ref{eq:eqtosolve}) basically is an equation for $R_{sub}$, with the boundary 
condition that $U(p_{max})=1$. Finding the value of $R_{sub}$ (and the 
corresponding value for $R_{tot}$) such that $U(p_{max})=1$ also provides 
the whole function $U(p)$ and, through eq. (\ref{eq:solut}), the distribution 
function $f_0(p)$ for the particles resulting from acceleration and 
reacceleration in the nonlinear regime. If the backreaction of the accelerated 
particles is small, the {\it test particle} solution must be recovered. 

\section{The nonlinearity of acceleration and reacceleration}
\label{sec:results}

In this section we investigate the shock modification due to the backreaction
of the accelerated particles. We split the discussion in two 
parts: in \S \ref{subsec:seed} we consider the case of pre-existing seed 
particles populating the region where the shock is propagating, and 
re-energized by the shock. 
We find the conditions for the onset of the nonlinear regime 
in which the shock gets modified by the re-accelerated particles. In particular
we show that the critical behaviour found for shock acceleration and 
manifesting itself through the appearance of three solutions, takes place 
also in this case.
In \S \ref{subsec:thermal} we explore the more complicated situation in 
which a shock accelerates a new population of particles while possibly 
reaccelerating a pre-existing population of seed particles.

\subsection{Shocks modified by re-energized seed particles}\label{subsec:seed}

Here we assume that a fluid is moving with speed $u_0$ in a region where the
temperature is $T_0$. The Mach number is some pre-defined value $M_0$, which 
can be easily related to $u_0$ and $T_0$. In the region of interest we assume
that a population of seed particles is present with energy per particle 
already higher than some injection energy necessary for the particles to 
{\it feel} the shock as a discontinuity. 
For simplicity we assume that these seed particles have a spectrum which 
is a power law in momentum in the form:
\begin{equation}
f_\infty(p)=f_{\infty,0} p^{-\alpha}.
\label{eq:f_infty}
\end{equation}
In this section we explore the situation in which the shock reacceleration of
a pre-existing population of cosmic rays can modify the shock structure, but
there is no acceleration of particles different 
from those that are already present as seed particles. This is equivalent to 
turn off the injection term ($\eta=0$ in Eq. (\ref{eq:laeffe})). In the next 
section we discuss what happens when both components are present.
The crucial difference between the two components is that the seed particles 
are by definition already above the threshold for {\it injection}, so that 
there is no injection efficiency that instead represents such a crucial 
parameter for the case of acceleration of particles extracted from the 
thermal pool.

Having in mind the case of shocks propagating in the interstellar medium of
our Galaxy, we consider here the case in which the inflowing gas and the 
seed particles are in pressure equilibrium, namely $P_{g,0}=P_{CR}$. In 
terms of the parameter $\xi_{CR}=P_{CR}/P_{g,0}$ this implies $\xi_{CR}=1$. 
The results can then be easily repeated for a generic value of $\xi_{CR}$.

Eq. (\ref{eq:eqtosolve}) gives the quantity $U(p)$ as a function of the 
momentum $p$ for any choice of $R_{sub}$ and $R_{tot}$, while the acceptable 
solutions are those with the right matching conditions at $p_{max}$. For 
simplicity, let us assume that the $p_{max}$ in eq. (\ref{eq:f_infty}) is 
the same maximum momentum that particles injected at the shock would achieve: 
this value only depends on the environmental conditions (energy losses of the 
particles) and/or on the geometry of the shock, which may allow the escape 
of the particles. The momentum $p_{max}$ is the same as in the distribution of 
the pre-exisiting seed particles if, for instance, the seed
particles have been accelerated by a shock identical to the one we are 
considering.
In any case, this assumption is not needed for the validity of our conclusions 
and may be easily relaxed, it simply serves to avoid parameter proliferation.
Within this assumption, particles re-energized at the shock are simply 
redistributed in the momentum range between $p_{inj}$ and $p_{max}$.

The solution, namely the right pair of values for the compression parameters 
$R_{sub}$ and $R_{tot}$ [related through eq. (\ref{eq:Rsub_Rtot})], is 
obtained when the solution corresponding to $U(p_{max})=1$ is selected.

\begin{figure}[thb]
 \begin{center}
  \mbox{\epsfig{file=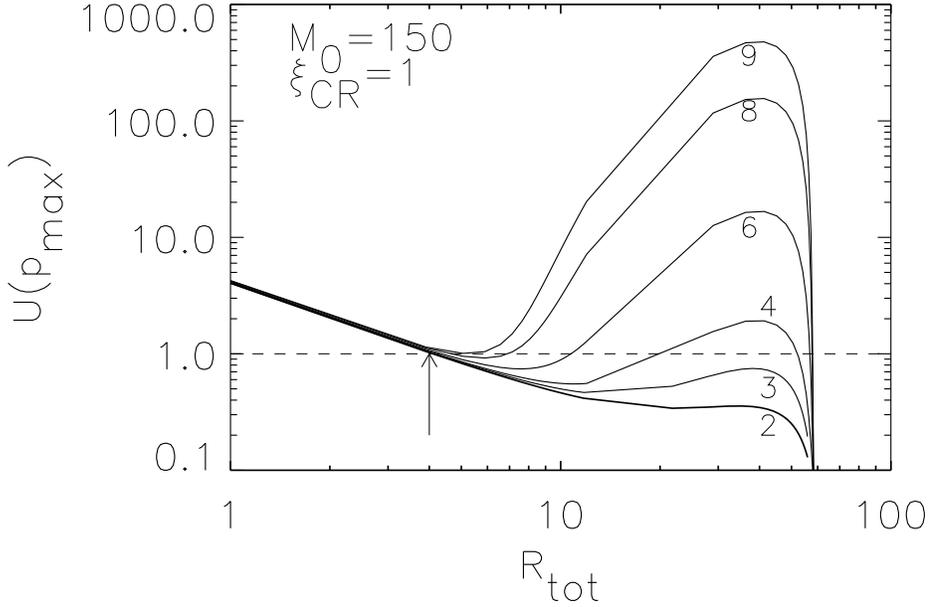,width=13.cm}}
  \caption{Velocity at upstream infinity in units of $u_0$ as obtained
from calculations, as a function of the total compression factor. The 
physical solutions are the ones with $U(p_{max})=1$.
The plot refers to the case $M_0=150$ and $\xi_{CR}=1$. The different lines 
are labelled by a number representing the $\log_{10}$ of $p_{max}$ in units 
of $m c$.}
\label{fig:M150}
 \end{center}
\end{figure}

In Fig. \ref{fig:M150}, we plot $U(p_{max})$ as a function of $R_{tot}$ for 
a shock having Mach number at infinity $M_0=150$. The different lines 
are labelled by a number representing the $\log_{10}$ of $p_{max}$ in units 
of $m c$. In other words the parameter $p_{max}$ changes between $10^2 m c$ 
and $10^9 m c$. The physical solutions are found by determining the 
intersections of each curve with the horizontal line corresponding to 
$U(p_{max})=1$. The minimum momentum of the seed particles is taken as 
$10^{-3}m c$.  
The intersection at $R_{tot}\simeq 4$ is close to the well known linear 
solution, namely the solution that one would obtain in the test particle 
approximation. 
The energy channelled into the nonthermal particles for this solution is very 
small, and the shock remains approximately unmodified. Increasing the value
of $p_{max}$, the solution moves toward slightly larger values of $R_{tot}$
(namely the shock becomes more modified). For some values of $p_{max}$,
multiple solutions appear. In particular three regions of $p_{max}$ can be 
identified: 

1) $p_{max}\leq p_{cr}^{(1)}$: in this region (low values of 
$p_{max}$) the only solution is very close to the one obtained in the 
test particle approximation. For the values of the parameters in 
Fig. \ref{fig:M150}, $p_{cr}^{(1)}\approx 10^3 m c$. 

2) $p_{cr}^{(1)}\leq p_{max} \leq p_{cr}^{(2)}$: in this region (intermediate
values of $p_{max}$ three solutions appear, two of which imply a strong 
modification of the shock, namely an appreciable part of the energy flowing 
through the shock is converted into energy of the accelerated particles. 
For the values of the parameters in Fig. \ref{fig:M150}, 
$p_{cr}^{(2)}\approx 10^9 m c$. 

3) $p_{max}\geq p_{cr}^{(2)}$: in this region (high values of $p_{max}$) the
solution becomes one again, and the shock is always strongly modified ($R_{tot}
\gg 4$). 

\begin{figure}[thb]
 \begin{center}
  \mbox{\epsfig{file=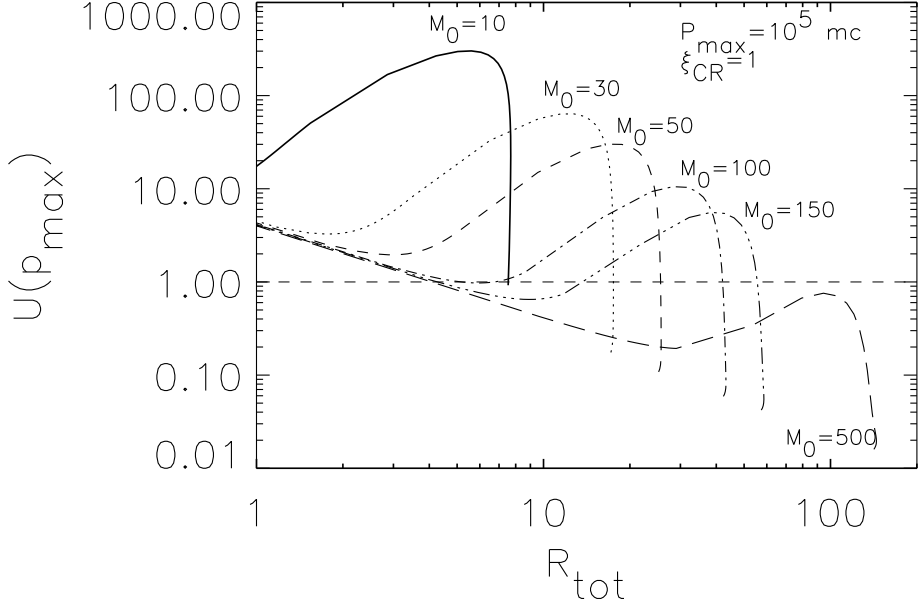,width=13.cm}}
  \caption{Velocity at upstream infinity in units of $u_0$ as obtained
from calculations as a funtion of the total compression factor. The physical 
solutions are the ones with $U(p_{max})=1$.
The plot refers to the case $p_{max}=10^5 m c$ and $\xi_{CR}=1$ for the Mach
numbers indicated on the curves. }
\label{fig:Pmax10_5}
 \end{center}
\end{figure}

\begin{figure}[thb]
 \begin{center}
  \mbox{\epsfig{file=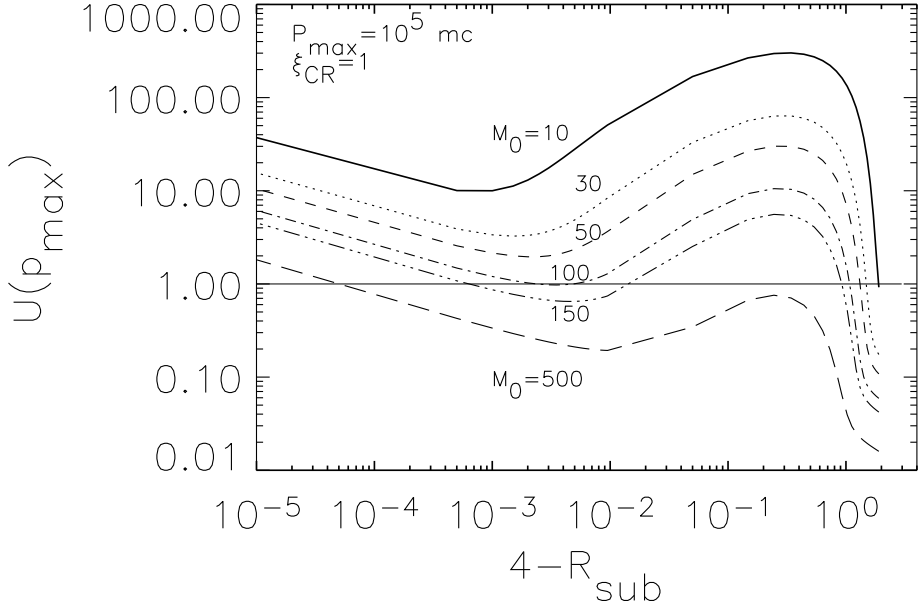,width=13.cm}}
  \caption{The same as in Fig. \ref{fig:Pmax10_5}, but as a function of 
$4-R_{sub}$.}
\label{fig:Pmax10_5RSUB}
 \end{center}
\end{figure}

This critical behaviour appears also when one fixes the maximum momentum 
and uses the Mach number of the shock as the order parameter. The results 
are shown in Fig. \ref{fig:Pmax10_5}, where Mach numbers between 10 and 
500 have been considered, at fixed $p_{max}=10^5 m c$. One can see that for 
Mach numbers below $\sim 100$ there is only one solution. For Mach numbers 
between $\sim 100$ and $\sim 500$ three solutions appear, one of which roughly 
corresponds to an unmodified shock. For larger Mach numbers, only
this linear solution remains, and the shock is always only weakly modified. 
The same situation is plotted in Fig. \ref{fig:Pmax10_5RSUB}, where instead of 
$R_{tot}$ on the x-axis there is $4-R_{sub}$, and $R_{sub}$ is the compression 
coefficient at the gas subshock. For unmodified shocks one expects 
$4-R_{sub}\to 0$. 
Again three regions can be identified, in the parameter $M_0$. 

1) For $M_0\leq M_{cr}^{(1)}$ only one solution exists and does not necessarily
correspond to unmodified shocks. In fact one can see that for 
$M_0=10<M_{cr}^{(1)}\sim 100$, the compression coefficient at the subshock is 
$\sim 2$, and $R_{tot}\sim 8$, for the situation plotted in Figs. 
\ref{fig:Pmax10_5} and \ref{fig:Pmax10_5RSUB}. Even
smaller values of $M_0$ do not give a perfectly unmodified shock. 

2) For $M_{cr}^{(1)}\leq M_0 \leq M_{cr}^{(2)}$ three solutions appear, two 
of which imply a strong modification of the shock. For the parameters used in 
Figs. \ref{fig:Pmax10_5} and \ref{fig:Pmax10_5RSUB}, $M_{cr}^{(2)}\sim 500$. 

3) For $M_0\geq M_{cr}^{(2)}$ only the nearly unmodified solution exists.

\begin{figure}[thb]
 \begin{center}
  \mbox{\epsfig{file=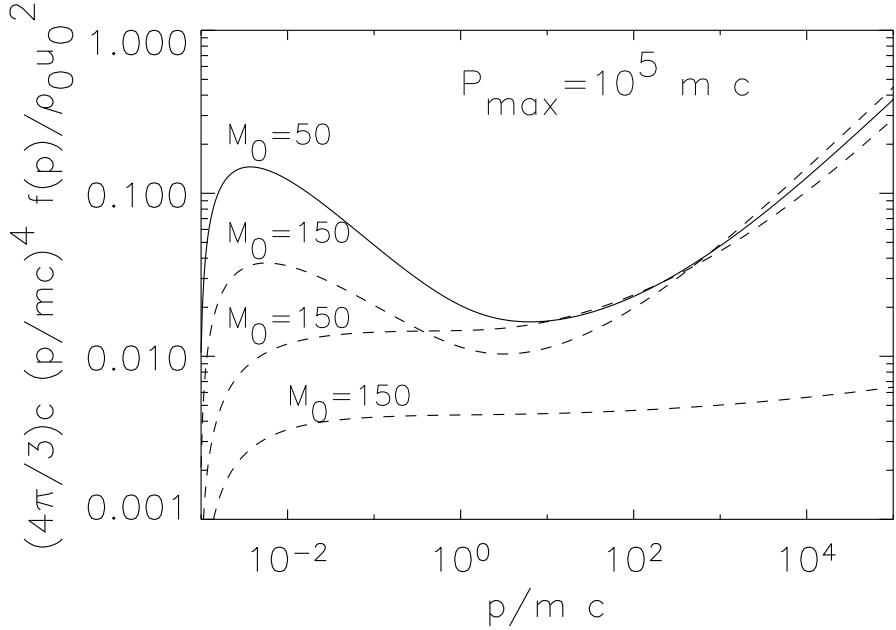,width=13.cm}}
  \caption{Spectra of re-accelerated particles for the cases $M_0=50$
(single solution) and $M_0=150$ (three solutions). The maximum momentum is
fixed to $p_{max}=10^5 m c$.}
\label{fig:spectra}
 \end{center}
\end{figure}

In order to emphasize the fact that the multiple solutions actually correspond 
to physical solutions with very different spectral characteristics for the 
accelerated particles we plot in Fig. \ref{fig:spectra} the spectra for 
$M_0=50$ (for which there is only one solution) and $M_0=150$ (for which 
there are three solutions). In both cases we used $P_{max}=10^5 m c$. 
The solution corresponding to Mach number $M_0=50$ has $R_{sub}=2.683$ and
$R_{tot}=25.588$, resulting in a total pressure of the accelerated particles
$P_{CR}/\rho_0 u_0^2=0.8946$ (solid line in Fig. \ref{fig:spectra}). The
three dashed lines in Fig. \ref{fig:spectra} are for $M_0=150$ and represent
the spectra for the three solutions. In numbers, these solutions (from top 
to bottom in Fig. \ref{fig:spectra}) are summarized as follows:\par\noindent
{\it first solution:} $R_{sub}=3.0775$, $R_{tot}=55.607$, 
$P_{CR}/\rho_0 u_0^2=0.9611$
\par\noindent
{\it second solution:} $R_{sub}=3.9839$, $R_{tot}=14.318$, 
$P_{CR}/\rho_0 u_0^2=0.734$
\par\noindent
{\it third solution:} $R_{sub}=3.999371$, $R_{tot}=4.2548$, 
$P_{CR}/\rho_0 u_0^2=0.06$

In the cases of strong shock modification, the asymptotic spectra for 
$(p/m c)\gg 1$ are $p^2 f(p) \sim p^{-3/2}$, while
the linear theory, in case of strong shocks ($M_0\gg 1$), would 
predict $p^2 f(p)\sim p^{-2}$. In the region of very low energies, the 
spectra of the reaccelerated particles tend to zero, as known from the 
linear theory as well.

\subsection{Modified shocks: the general case}\label{subsec:thermal}

Nonlinear effects in shock acceleration were first investigated in the 
case when a fraction of the thermal gas crossing the shock surface is 
energized to nonthermal energies. In this section we wish to apply the 
analytical method discussed in \S \ref{sec:nonlin} and \S \ref{sec:gasdyn} 
to the most general case in which thermal particles are accelerated but 
seed particles may already be present in the environment.

\begin{figure}[thb]
 \begin{center}
  \mbox{\epsfig{file=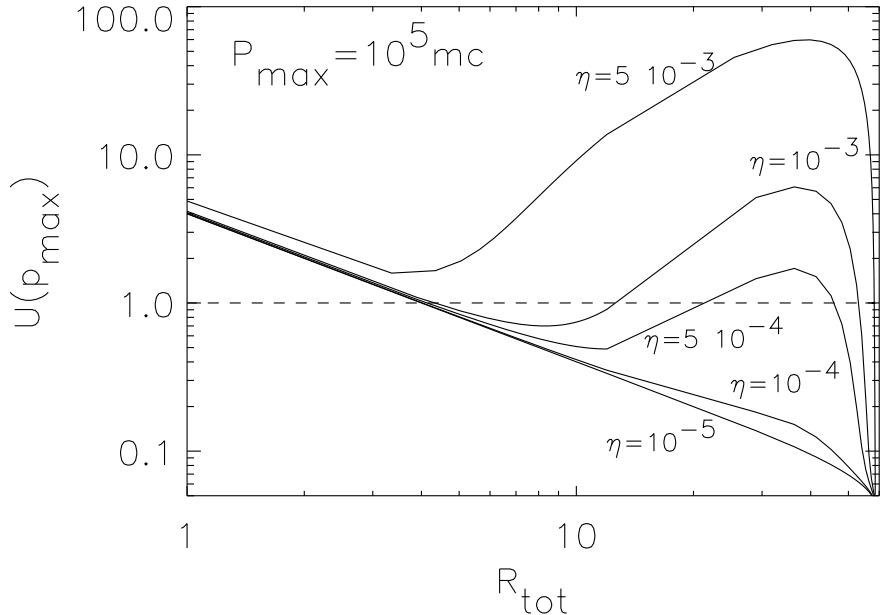,width=13.cm}}
  \caption{Velocity at upstream infinity in units of $u_0$ as obtained
from calculations, as a function of the total compression factor. Only 
particles injected at the shock from the thermal pool are considered here
with efficiency $\eta$ indicated on the curves.}
\label{fig:NOseed}
 \end{center}
\end{figure}

\begin{figure}[thb]
 \begin{center}
  \mbox{\epsfig{file=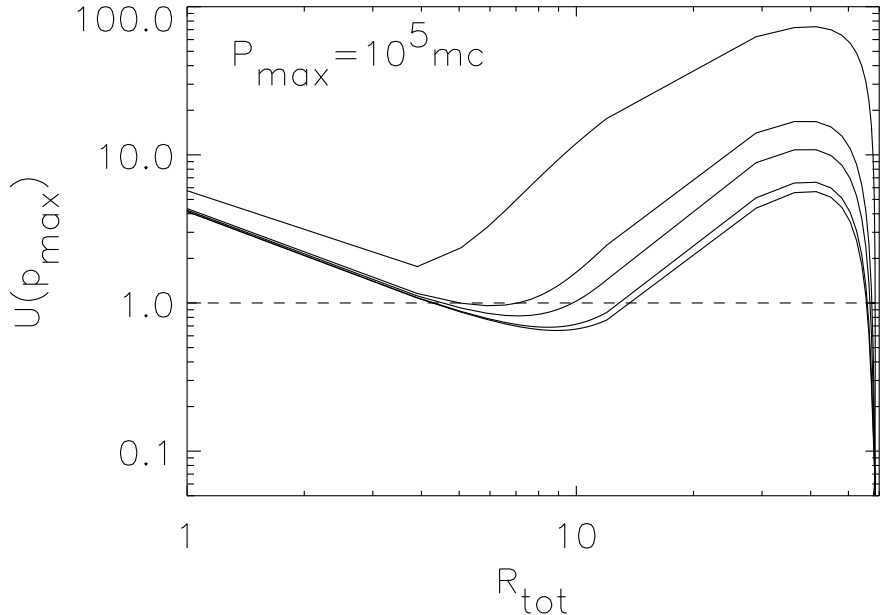,width=13.cm}}
  \caption{Velocity at upstream infinity in units of $u_0$ as obtained
from calculations, as a function of the total compression factor. Both 
particles injected at the shock from the thermal pool and seed particles
in pressure equilibrium with the upstream infinity gas pressure are 
considered here. The efficiencies are as in Fig. \ref{fig:NOseed}.}
\label{fig:CRseed}
 \end{center}
\end{figure}

We start our discussion with the case of acceleration of particles from the 
thermal distribution, in order to show that the multiple solutions already 
found in other analytical approaches \cite{malkov1,malkov2} are also obtained 
by adopting the approach illustrated here. In Fig. \ref{fig:NOseed}, we plot 
the $U(p_{max})$ as a function of $R_{tot}$ for the case in which a fraction 
$\eta$ of the particles (as indicated in the plots) crossing the shock is 
actually accelerated to suprathermal energies. The Mach number is chosen 
to be $M_0=150$ and the maximum momentum is taken as $P_{max}=10^5 mc$.
One can easily see that $R_{tot}\sim 4$, the value for unmodified shocks, for 
small values of $\eta$ (low efficiencies), while $R_{tot}$ increases above 4 
for $\eta>10^{-4}$. Three solutions appear for intermediate efficiencies, while
the solution for the accelerated particles always predicts a strongly 
modified shock for high efficiencies, $\eta>3\times 10^{-3}$. An asymptotic
value of $R_{tot}\sim 60$ is achieved for the parameters used here. 
This example demonstrates that the critical behaviour shown to appear for high
values of $\eta$ in several analytical or semi-analytical calculations is in 
fact also predicted by the approach proposed here. 
Next question to ask is however whether the presence of seed particles can 
change the critical behavior of shocks. In order to answer this question we 
consider a situation in which particles are accelerated from the thermal pool 
with efficiency $\eta$, and at the same time the shock propagates in a medium 
where the preshock pressure in seed particles equals the gas pressure. This 
situation is considered to resemble that of a supernova explosion in our 
Galaxy, where cosmic rays fill the volume remaining in quasi-equipartition 
with the gas. In order to test the critical structure of the shock we 
calculate $U(p_{max})$ as a function of the compression factor $R_{tot}$ 
between upstream infinity and downstream. Our results are plotted in Fig. 
\ref{fig:CRseed}. The lines refer to the cases $\eta=10^{-5},~10^{-4},
~5\times 10^{-4},~10^{-3},~5\times 10^{-3}$ from bottom to top, as in Fig.
\ref{fig:NOseed}.
It is clear from this plot that the shock may be modified by the backreaction 
of the accelerated particles even for very low values of $\eta$, because of 
the presence of seed particles. In other words, the nonlinearity of the shock 
can well be dominated by the presence of preaccelerated particles rather than 
by the acceleration of particles from the thermal pool. Clearly this depends 
however on the values of the parameters (injection momentum, $\eta$, maximum 
momentum achievable, Mach number). The injection momentum, in principle, could 
be related to $\eta$ in order to reduce the number of free parameters of the 
problem. This is possible in the assumption that the particles downstream 
keep a thermal distribution. Simulations suggest that the particles
injected in the accelerator are the ones with momentum a few times the thermal 
momentum of the particles downstream, so that $\eta$ can be simply calculated 
by integration of the Maxwell-Boltzmann (MB) distribution above this minimum 
momentum. Unfortunately, we do not know whether the particle distribution 
downstream is in fact Maxwell-Boltzmann-like. Moreover, we do not know exactly 
what is the threshold to impose on the injection momentum, which is a delicate 
issue because the number of particles taking part to the acceleration process 
would result from the integral of the MB distribution over its exponentially 
decreasing part. Therefore we preferred here to keep $\eta$ and the injection 
momentum as separate free parameters.

\section{Discussion and Conclusions}\label{sec:conclusions}

We proposed a semi-analytical approach to show that the backreaction of 
particles accelerated at a shock is able to affect the shock itself, in 
such a way that the shock and the accelerated particles become parts of 
a nonlinear system. In particular, for the first time we included in this 
kind of calculations the seed particles that may be present in the region 
where the shock is propagating and that can be re-energized by the shock. 

While a test-particle approach to this problem was first presented in 
\cite{bell78}, a nonlinear treatment was never investigated. In the 
pioneering work of Ref. \cite{bell78}, it was recognized that the energy 
of the seed particles could be enhanced by about one order of magnitude 
at each shock passage, and that after an infinite number of strong shocks 
passing through the region, the spectrum of the particles would tend to 
the asymptotic spectrum $E^{-3/2}$. Two comments are in order. First, 
the continuous increase of the cosmic ray energy due to the re-energization 
of seed particles leads unavoidably the shock to be modified by 
the nonthermal pressure, unless one starts from an uninterestingly 
small pressure of seed particles at the beginning. Second, the fact that 
the spectrum becomes flatter than $E^{-2}$, which is the result for a 
strong non-relativistic shock, implies that most of the nonthermal energy 
is pushed to the highest energies, therefore the shock again can be more 
easily modified. Both these points suggest that a nonlinear treatment of 
shock re-acceleration is required. 

In our Galaxy, cosmic rays are observed to be in rough equipartition with the
gas pressure and with magnetic fields, therefore supernova shocks or shocks 
generated in other environments propagate in a medium in which the seed 
particles (cosmic rays) are non-negligible. In these circumstances the 
non-linear effects may be very important. We showed here that in fact for 
some regions of the parameter space, the shock is modified mainly by the 
backreaction of the seed particles rather than by the cosmic rays accelerated 
at the shock from the thermal pool. The spectra of the re-accelerated 
particles have also been calculated. 

The interesting phenomenon of the appearance of multiple shock solutions, 
already known for the case of shock acceleration, appears also for the case 
of reacceleration. This puzzling phenomenon may suggest that the shock
behaves as a self-regulating system settling on the critical point, as 
proposed in \cite{malkov1,malkov2}. On the other hand, it is possible that
the multiple solutions may be the artifact of some of the assumptions 
used in the analytical approach, in particular the request for time 
independent (stationary) solutions and the fact that the role of the 
self-generated waves on the diffusion coefficient is not taken into 
account. Further investigation, in particular in the direction of 
a detailed comparison of our results with numerical simulations of
shock acceleration is required in order to unveil the physical 
meaning of the multiple solutions for modified shocks.

From the phenomenological point of view, it would be certainly worth
to study the implications of nonlinear shock reacceleration on the
nonthermal activity in astrophysical environments where the effect is
expected to play an important role, in particular in the case of supernova 
remnants. In particular, as pointed out by the referee, the suggested 
dominance of reaccelerated ambient seed particles over freshly injected 
particles may have serious implications for the spectra of secondary 
nuclei resulting from spallation processes.

\section*{Acknowledgments} The author gratefully acknowledges useful
discussions with M. Baring, D. Ellison and M. Vietri. The author is also
grateful to the referee, L.O'C. Drury for his very useful remarks. This 
work was partially supported through grant COFIN 2002 at the Arcetri 
Astrophysical Observatory.


\end{document}